\documentclass[runningheads]{llncs}
\usepackage[T1]{fontenc}
\usepackage{graphicx}
\usepackage{hyperref}
\usepackage{color}

\begin{document}
\title{Evaluation of Computational and Power Performance in Matrix Multiplication Methods and Libraries on CPU and GPU using MKL, cuBLAS, and SYCL}
%
%
\author{L.A. Torres\inst{1,2,3}\orcidID{0000-0003-2597-9430} \and
Carlos J. Barrios H.\inst{1,2,3}\orcidID{0000-0002-3227-8651} \and 
Yves Denneulin \inst{4,5}\orcidID{0000-0002-0340-2094}}
\authorrunning{L.A. Torres Carlos J. Barrios H., Y. Denneulin}
\institute{Supercomputación y Cálculo Científico (SC3UIS) \and
Grupo de Investigación Computo Avanzado y a Gran Escala (CAGE) \and
Universidad Industrial de Santander, Bucaramanga 680002, Colombia \and
Universit\'{e} Grenoble Alpes, Grenoble, France \and
Laboratoire d'Informatique de Grenoble (LIG), Grenoble, France
\email{luis.torres@correo.uis.edu.co}\\
\email{cbarrios@uis.edu.co}
\email{yves.denneulin@@grenoble-inp.f}
}
\maketitle              
%
\begin{abstract}
Matrix multiplication is fundamental in the backpropagation algorithm used to train deep neural network models. Libraries like Intel's MKL or NVIDIA's cuBLAS implemented new and optimized matrix multiplication techniques that increase performance and reduce computational costs. These techniques can also be implemented in CUDA and SYCL and functions with AVX2 and AVX512 instructions, which have lower performance but better precision. The study compares execution times and power consumption using PAPI and PERF and compares accuracy for different matrix sizes. Comparisons were made on architectures such as third and fourth-generation Intel CPUs and NVIDIA V100 and A100 GPUs. The MKL library showed the best performance with a slight loss of precision, while OpenMP and SYCL on the CPU implementation showed the best accuracy but a loss of performance.
On the other hand, the results on GPU showed that cuBLAS with tensor cores had the best performance; however, it had a cost in accuracy. The cuBLAS library without these specialized cores shows minimal performance loss and much higher accuracy. The data obtained on different architectures showed that the CPU could achieve performance close to that obtained on the GPU with increased power consumption. These results are conditional on certain hardware specifications, such as the number of cores, clock frequency, processor generation for the CPU, and the speed and bandwidth of the PCI bus and device architecture (compute capability) for the GPU.

\end{abstract}
%
%
\section{Introduction}
In recent years, heterogeneous computing systems have become very popular because some parts of the application code work better on hardware other than the CPU. These systems have become one of the solutions concerning the power efficiency required by current policies. Fields such as computational fluid dynamics and machine learning have increased their computational load, requiring more computing power. This increase makes it necessary to orchestrate these heterogeneous systems while balancing performance and power consumption efficiently. \cite{Cussen2023matrix}. 

Matrix multiplication is a fundamental operation in many fields of science and engineering. In deep learning, optimizing this operation is a field of new interest because it consumes a significant fraction of the model training time. The main reason is that the backpropagation algorithm \cite{Rumelhart1986bp} is a matrix multiplication sequence. Since the 1950s, there has been a search for a NxN matrix multiplication algorithm that reduces the time the ordinary algorithm takes of order $O(n^3)$, like Strassen \cite{Strassen1969Gaussian} with $O(n^{2.81})$ to the best known today with $O(n^{2.37})$ \cite{Alman2021FasterMM}. In high-performance computing, these algorithms are grouped into libraries called GEMM (general matrix multiplications) routines.

Concerning deep neural network workloads, typical matrix sizes in GEMM operations are close to 10.000\cite{Qin2020SparceGEMM},  so it is required to split arrays effectively to fit into the memory hierarchy and improve performance. A layered approach is used in current numerical libraries \cite{Goto2008AnatomyMM}, reorganizing data to improve locality as it moves from the main memory through the memory hierarchy \cite{Kuzma2023FastMM}. Different libraries use this layered approach in a highly optimized way. However, they all have drawbacks, such as the requirement that each library be installed according to the architecture, each library must be written in the assembly code of each architecture, and gaps between the release of new architectures and the creation of libraries optimized for said architecture.

For decades, the CPU was the leading architecture in processing scientific calculations, which motivated the algorithm designs to focus on being able to execute these routines in parallel on a large number of cores CPU. In recent years, with the apparition of GPUs, these designs were adapted to these new architectures, which have shown better performance for these highly parallel tasks \cite{Baratta2022PerfMat}. Intel and NVIDIA provide highly optimized libraries to run these routines on their respective architectures, Intel oneAPI Math Kernel Library (MKL) and CUDA Basic Linear Algebra Subroutine (cuBLAS). Works such as those of Khalilov and Timoveev \cite{Khalilov2021PerfCUDA} and Krainiuk et al. \cite{Krainiuk2021oneAPI} have evaluated the performance of these libraries with simple matrix-vector multiplications and other simple algorithms, showing excellent efficiency and performance in their calculations.

NVIDIA introduced a specialized unit called Tensor Core with its Volta microarchitecture, which manages to perform a $4x4$ matrix multiplication per clock cycle. The first versions of this architecture provided 640 Tensor Cores, achieving a theoretical performance of 125 Tflops/s in mixed precision but incurring a loss of precision, which can be critical in many HPC applications. However, it considerably reduces execution time \cite{Yan2020DemysTensor}. Currently, NVIDIA offers three ways to perform matrix multiplication and accumulation (FMA - fused multiply-add) using Tensor Cores: CUDA Warp Matrix Multiply Accumulate (WMMA) API, CUTLASS - WMMA-based template library and cuBLAS GEMM \cite{Markidis2018TensorProg}. 

The growing heterogeneity in HPC architectures has led to the search for new ways to reduce the programming complexity and the learning curve of the several dozen existing libraries and standards. In 2009, OpenCL\footnote{https://www.khronos.org/opencl/} appeared as the first development framework to support this architecture type. Its main drawback was due to the very low-level programming that it required. In 2014, SYCL\footnote{https://www.khronos.org/sycl/} appeared, an open programming model that, like OpenCL, allows the programming of heterogeneous systems, presenting great portability without the complexity of its predecessor. The SYCL runtime implicitly considers low-level details, such as memory management and data synchronization. Its main feature is that it uses the SMPC (single source multiple compiler-passes) approach to compile both the host and device codes, generating a thick executable file that can be executed on both architectures \cite{Reddy2021PergGPU} \cite{daSilva2016CompSYCL}.

This work performs a comparative study of the matrix multiplication algorithm on different architectures, measuring execution times and power consumption. It also seeks to determine the possibility of executing a distributed algorithm for training neural networks by distributing workloads (matrix multiplications during forward and backward propagation). The evaluation compared matrix multiplication techniques, including openMP, intrinsic functions (AVX2 and AVX512), and CUDA. It also included Intel's MKL and NVIDIA's cuBLAS libraries. For CUDA and CUBLAS, tests were done with and without Tensor Cores. Finally, this study evaluated SyCL on CPU and GPU and compared the performance and power consumption with the other techniques and libraries mentioned above.

This article is organized as follows: first, an introduction with different works on the performance of matrix multiplication libraries, SYCL performance evaluations, and evaluations of specific hardware characteristics, particularly in NVIDIA accelerators. Section II presents the methodology; Section III presents the results obtained. Finally, in the last section, the conclusions of this study.

\section{Methodology}
\subsection{Matrix Multiplication}
Many matrix multiplication methods have been adapted and optimized for computational environments. These methods are designed considering the organization of the different memory levels and the hardware architecture where they are programmed and executed. The simple swapping of a for loop in the matrix multiplication algorithm reduces the matrix multiplication time by reducing the number of cache misses. This method and several others help reduce execution times even further.

The tests addressed in this research implemented square matrix multiplication techniques using CUDA for NVIDIA GPUs, intrinsic functions with AVX2 and AVX512 processor instructions on CPUs, openMP, and SYCL. The tests over SYCL are carried out to evaluate its portability (main characteristic), performance, and power consumption. The data type of the matrices is fp32.

Finally, the libraries implemented were NVIDIA's cuBLAS and Intel's MKL. These libraries optimize this type of operation to obtain the highest possible performance for specific architectures. The study contrasted the libraries' results with those obtained with the implementations over CUDA, OpenMP, SYCL, and intrinsic functions, evaluating processing time and power consumption. It is important to note that each NxN matrix multiplication was executed ten times and averaged to look for stable results when comparing execution times. For power consumption, these matrix multiplications were executed twenty times. The number of repetitions in power consumption increased because, during several tests, it was observed that PAPI did not record values for the required metrics. Matrix sizes were set from 32x32 to 8192x8192. This margin was established to study the behavior of the implementations for small matrices up to matrices sizes that could represent similar numbers of GEMM operations in deep neural network workloads.

\subsection{Experimental Environment}
This study used three platforms with different architectures to evaluate time and power consumption. In the first platform, three types of servers with different hardware configurations were chosen from the grid5k infrastructure\footnote{https://www.grid5000.fr} (Table \ref{hardware1} and Table \ref{hardware2}). A custom image was used for Kadeploy\footnote{https://kadeploy.imag.fr/} with the Rocky Linux Version 9.2 operating system and the Linux 5.14.0 kernel.

The second server platform is the PACCA of the University of Cartagena (Colombia) cluster, which has more recent processors and GPUs. The hardware is described in Tables \ref{hardware3} and \ref{hardware4}.

Final testing was performed on Intel Developer Cloud with 4th-generation Intel processors. As for PAPI, the hardware performance counter necessary to evaluate the power consumption could not be configured, and therefore, results could only be obtained using PERF. Table \ref{hardware5} shows the server's hardware configuration.

\begin{table}
\caption{GRID5K Hardware Description.}\label{hardware1}
\begin{tabular}{|l|l|l|l|l|l|l|l|}
\hline
\textbf{\#} & \textbf{CPU Type} & \textbf{CPU} & \textbf{Cores} & \textbf{Freq(GHz)} & \textbf{Mem(GiB)} & \textbf{Freq(MHz)} & \textbf{PCI} \\
\hline
1 & Intel Xeon Gold 6126 & 2 & 48 & 2.60 & 192 & 2666 & 3.0 \\
2 & Intel Xeon Gold 6254 & 2 & 72 & 3.10 & 384 & 2933 & 3.0 \\
3 & Intel Xeon Silver 4314 & 2 & 64 & 2.40 & 256 & 3200 & 4.0\\
\hline
\end{tabular}
\end{table}

\begin{table}
\caption{GPU description for server No. 1.}\label{hardware2}
\begin{tabular}{|l|l|l|l|l|l|l|}
\hline
\textbf{GPU Type} & \textbf{Cores} & \textbf{T. Cores} & \textbf{freq(MHz)} & \textbf{Mem(GB)} &  \textbf{Freq(MHz)} \\
\hline
NVIDIA Tesla V100 & 5120 & 640 & 1230 & 32 & 876 \\
\hline
\end{tabular}
\end{table}

\begin{table}[h!]
\caption{PACCA Hardware Description.}\label{hardware3}
\begin{tabular}{|l|l|l|l|l|l|l|l|}
\hline
\textbf{\#} & \textbf{CPU Type} & \textbf{CPU} & \textbf{Cores} & \textbf{Freq(GHz)} & \textbf{Mem(GiB)} & \textbf{Freq(MHz)} & \textbf{PCI} \\
\hline
4 & Intel Xeon Gold 5320 & 2 & 104 & 2.20 & 256 & 3200 & 4.0 \\
5 & Intel Xeon Gold 5315Y & 2 & 32 & 3.20 & 256 & 3200 & 4.0 \\
\hline
\end{tabular}
\end{table}

\begin{table}
\caption{GPU description for server No. 5.}\label{hardware4}
\begin{tabular}{|l|l|l|l|l|l|l|}
\hline
\textbf{GPU Type} & \textbf{Cores} & \textbf{T. Cores} & \textbf{freq(MHz)} & \textbf{Mem(GB)} &  \textbf{Freq(MHz)} \\
\hline
NVIDIA A100 & 6912 & 432 & 765 & 40 & 1215 \\
\hline
\end{tabular}
\end{table}

The software used for the compilations was CUDA and cuBLAS 12.4 and oneAPI 2023, configured with the plugin\footnote{https://developer.codeplay.com/products/oneapi/nvidia/home/} that allows running on NVIDIA GPUs. The latter was used to compile the tests with SYCL. Regarding the energy consumption tests, PAPI (Performance Application Programming Interface) in version 7.1.0 was used with a complete installation, in which each of the hardware counters available in the treated architectures was activated except for the server in Intel Cloud, as mentioned before, no hardware counters were found that allowed power consumption to be measured in a similar way to the other servers. The version of PERF used was 4.18. The power consumption in both PAPI and PERF includes processor and memory consumption. The RAPL and NVML events corresponding to energy consumption were used in PAPI.

\begin{table}[h!]
\caption{Intel Cloud Hardware Description.}\label{hardware5}
\begin{tabular}{|l|l|l|l|l|l|l|l|}
\hline
\textbf{\#} & \textbf{CPU Type} & \textbf{CPU} & \textbf{Cores} & \textbf{Freq(GHz)} & \textbf{Mem(GiB)} & \textbf{Freq(MHz)} & \textbf{PCI} \\
\hline
6 & Xeon Platinum 8480+ & 2 & 224 & 2.0 & 512 & 4800 & 5.0 \\
\hline
\end{tabular}
\end{table}

The deployment technology provided by Grid5k, known as Kadeploy, allowed the deployed operating system image to be the same on each server. The Kadeploy image maintained the same configurations and versions across all deployed tests. Identical software versions were installed natively on the other servers.

\section{Evaluation and Results}
The results of the study have been divided into two sections. The first shows the execution times, while the second presents the power consumption. The execution times were taken exclusively from the function responsible for multiplying NxN matrices, except for the CUDA and cuBLAS versions, where the data transfer times between the host and device memory were also considered. Concerning power consumption, the data were measured with both PAPI and PERF. However, it should be clarified that the results presented with PAPI only include the power consumption of the function responsible for the matrix multiplication process. In contrast, PERF measures the total consumption of the code executed, including data load functions.
\subsection{Performance and Precision}
The matrix multiplication methods and libraries were evaluated on six processors from 2 different generations of Intel. The object of the study is mainly MKL and SYCL on CPU and cuBLAS and SYCL on GPU. However, tests were performed using OpenMP and intrinsic functions with AVX2 and AVX512 instructions to determine whether these are options for optimizing the matrix multiplication process. The graphs in this section present the time results on the left side in milliseconds. On the right side, it shows the mean square error (MSE) between the result obtained from the library or method evaluated and the result obtained from a serial function.

\begin{figure}
\includegraphics[width=\textwidth]{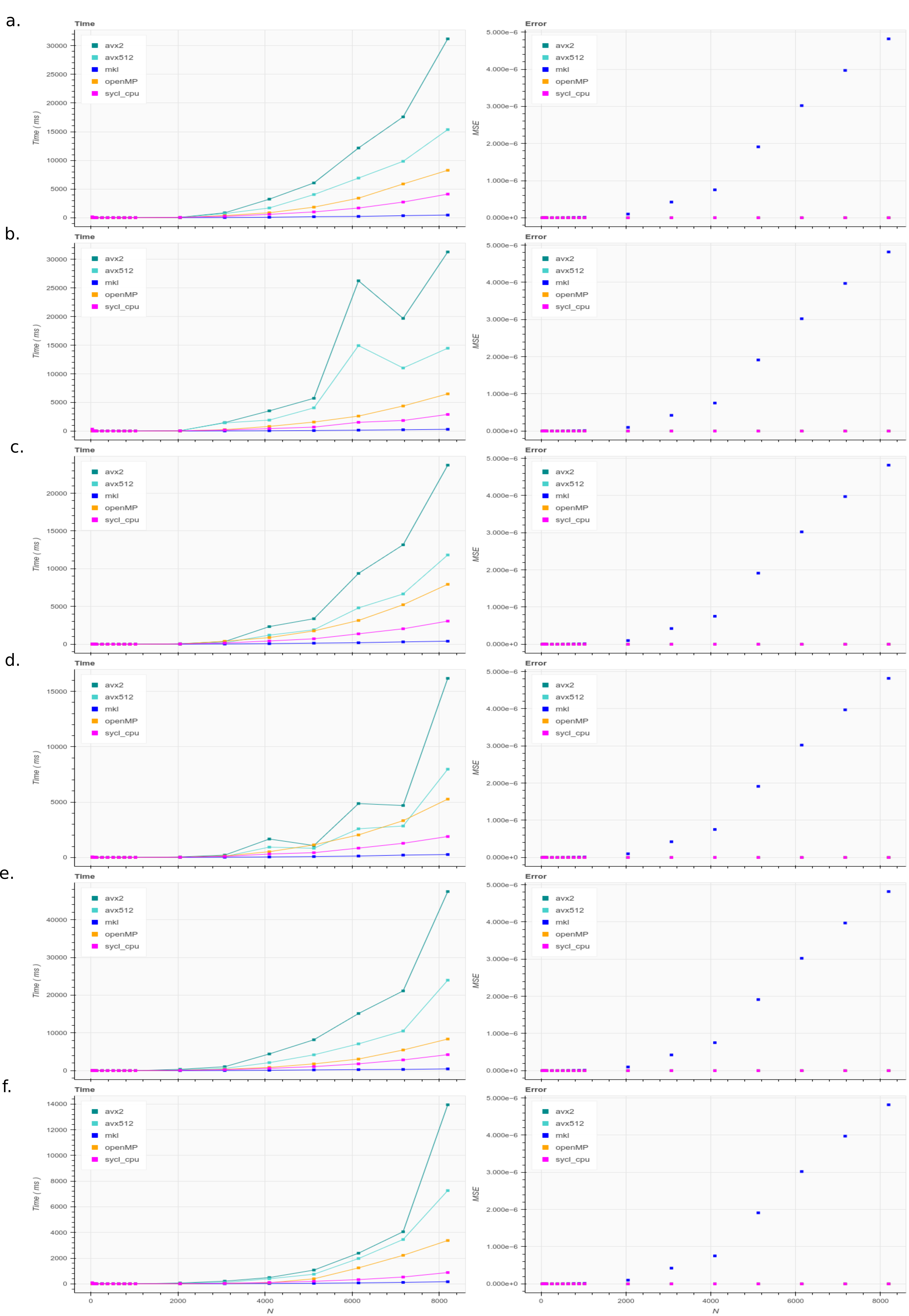}
\caption{Comparison of CPU execution times and MSE. a. Intel Xeon Gold 6126 (48 cores @ 2.60 GHz) b. Intel Xeon Gold 6254 (72 cores @ 3.10 GHz c. Intel Xeon Silver 4314(64 cores @ 2.40 GHz) d. Intel Xeon Gold 5320 (104 cores @ 2.20 GHz) e. Intel Xeon Gold 5315Y(32 cores @ 3.20 GHz) f. Intel Xeon Platinum 8480+ (224 cores @ 2.0 GHz)} \label{fig1}
\end{figure}

Figure \ref{fig1} shows the data obtained for the different processors evaluated. Regardless of the version or generation of the Intel processor, the timing results of AVX2 and AVX512 were the worst. It is possible to optimize our matrix multiplication functions that used intrinsic functions with prefetching and several other techniques. However, the code with intrinsic functions is quite complex, so it has been discarded as an option.

The best results were obtained from MKL and SYCL. OpenMP takes at least twice as long as SYCL, so it is discarded. When observing the error results of MKL and SYCL exclusively, an increasing error is observed in MKL. However, this is of the order of $10^{-6}$.

\begin{figure}[h!]
\includegraphics[width=\textwidth]{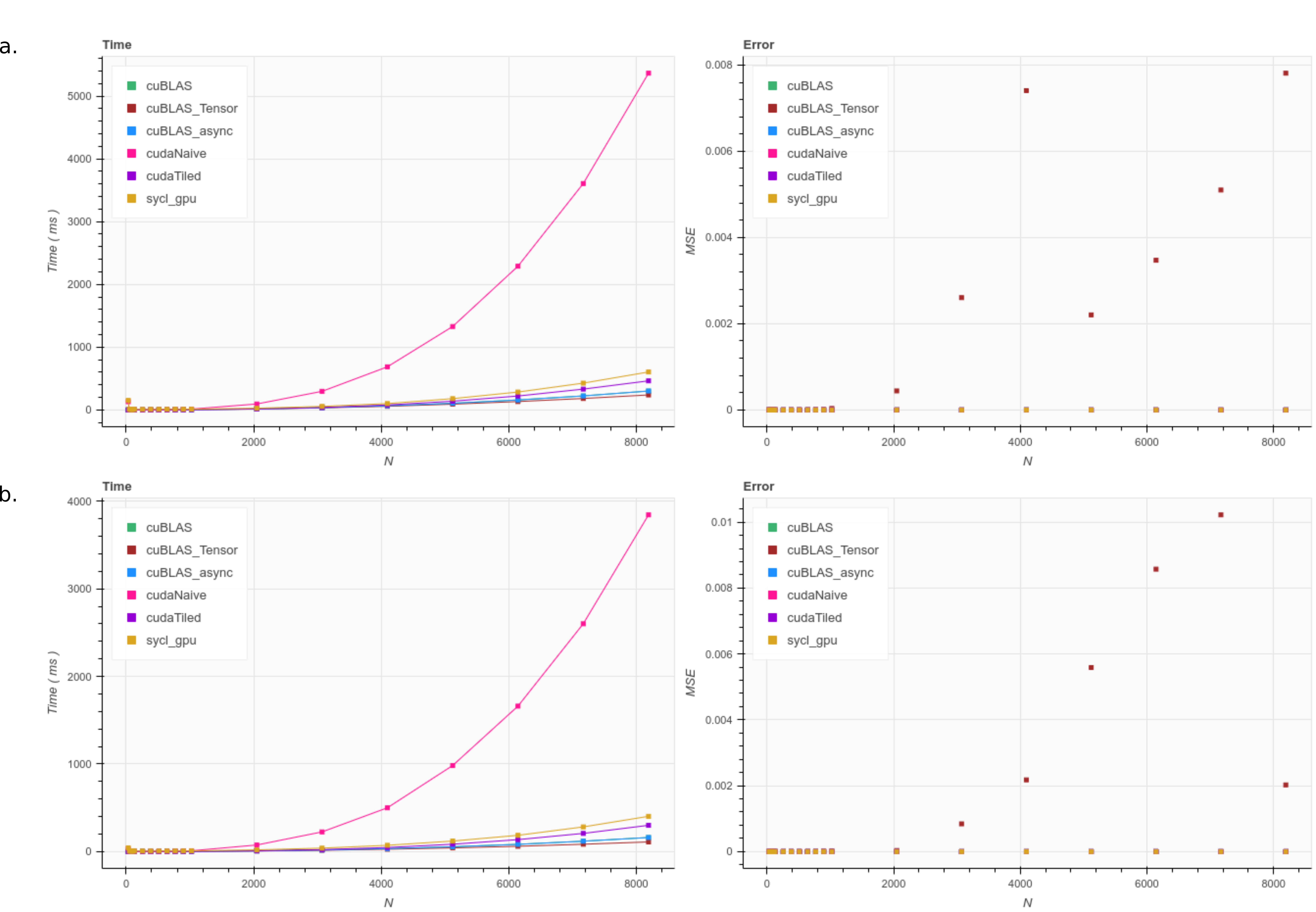}
\caption{Comparison of the execution times of the two NVIDIA architectures evaluated. a. Intel Xeon Gold 6126 (48 cores @ 2.60 GHz) - Tesla V100-PCIE-32GB b. Intel Xeon Gold 5315Y(32 cores @ 3.20 GHz) - NVIDIA A100-PCIE-40GB} \label{fig2}
\end{figure}

Figure \ref{fig2} shows the results of the two NVIDIA architectures evaluated. Different functions were built using CUDA code to determine whether they were viable. The code without optimizations obtained the worst result; However, when using the Tiled Matrix Multiplication method with CUDA, better times were achieved, even improving the SYCL times. It should be noted that the SYCL function was not optimized for any architecture; the lambda function was constant for CPU and GPU. The best results, as expected, were those of the cuBLAS library with or without Tensor Cores. However, the MSE presented using Tensor Cores is considerably higher than without using these, particularly in large matrices.

The results showed that MKL for Intel processors and cuBLAS for NVIDIA GPUs showed the best times. However, SYCL on the CPU showed acceptable performance without loss of precision, as did the optimized CUDA function on the GPU.

\begin{figure}[h!]
\includegraphics[width=\textwidth]{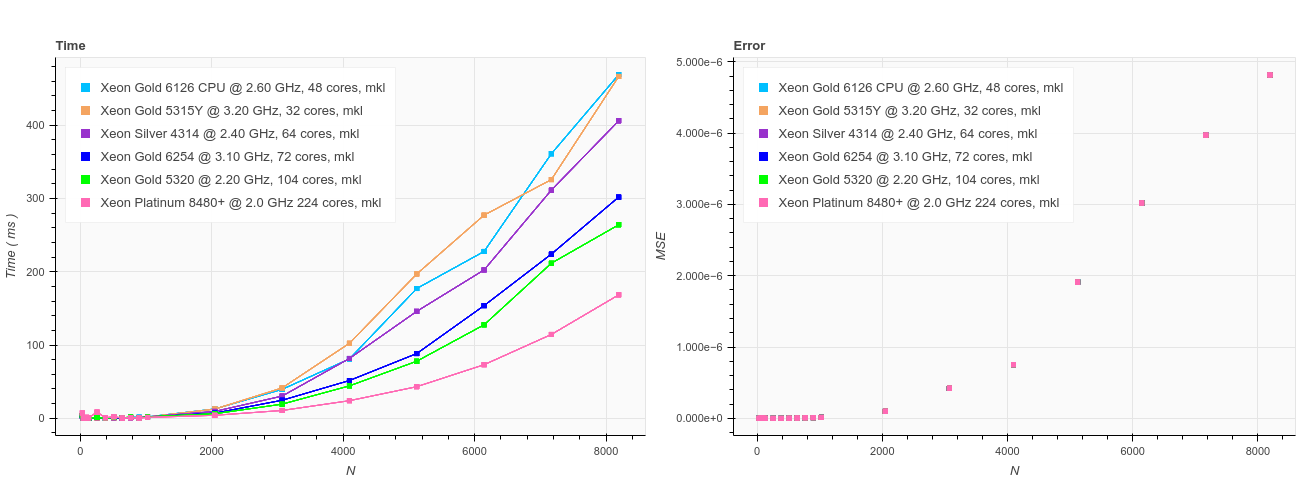}
\caption{Comparison of the execution times of the different processors evaluated using the MKL library.} \label{fig3}
\end{figure}

Figure \ref{fig3} compares each of the processors studied. This comparison was made with MKL, which presented the best time results. As mentioned, the MSE presented by MKL is of the order of $10^{-6}$. Notwithstanding the change in processor type and generation, it presents the same ascending behavior, and the error presents a slight variation between architectures. The processors that recorded the best times were the fourth-generation Intel Xeon Platinum 8480+ with 112 cores at 2.0 GHz each. The 104 cores of the two third-generation Intel Xeon Gold 5320 2.2 GHz processors came in second place. It is essential to note the relationship between cores and frequency regarding processing times. An increase in frequency allows fewer cores to achieve performance similar to processors with many cores at low frequencies. This situation can be observed in the Intel Xeon Gold 6254 and Intel Xeon Gold 5320 processors, where the 72 cores at 3.10 GHz of the former are currently very close to the 104 cores at 2.2 GHz of the latter.

\begin{figure}[h!]
\includegraphics[width=\textwidth]{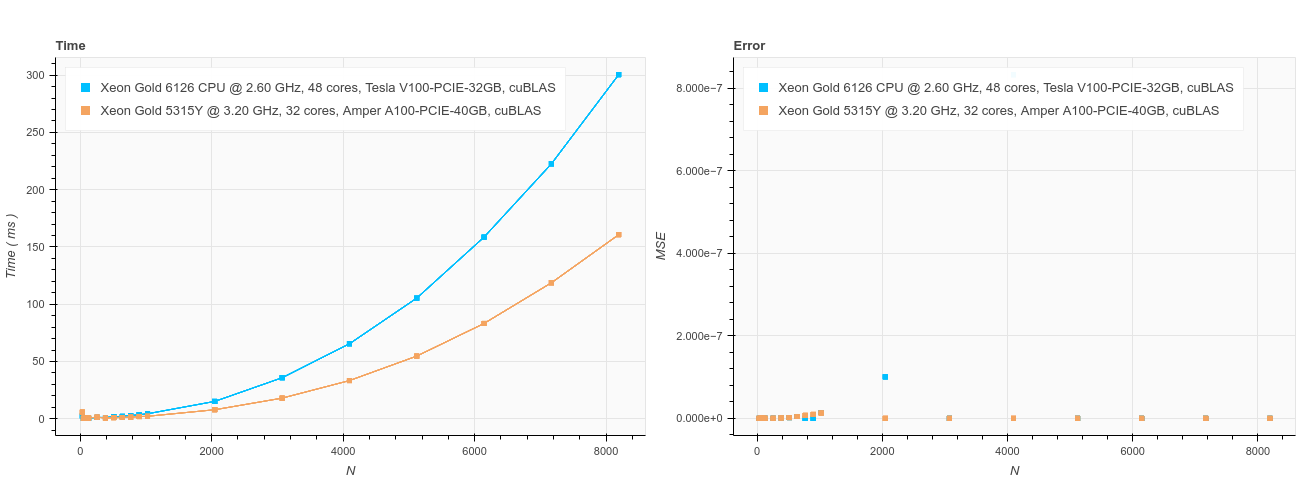}
\caption{Comparison of the execution times of the different GPUs evaluated using the cuBLAS library.} \label{fig4}
\end{figure}

Figure \ref{fig4} shows the times obtained from the two GPU architectures evaluated. NVIDIA A100 presented the best processing times and lower MSE than its predecessor NVIDIA V100.

\begin{figure}[h!]
\includegraphics[width=\textwidth]{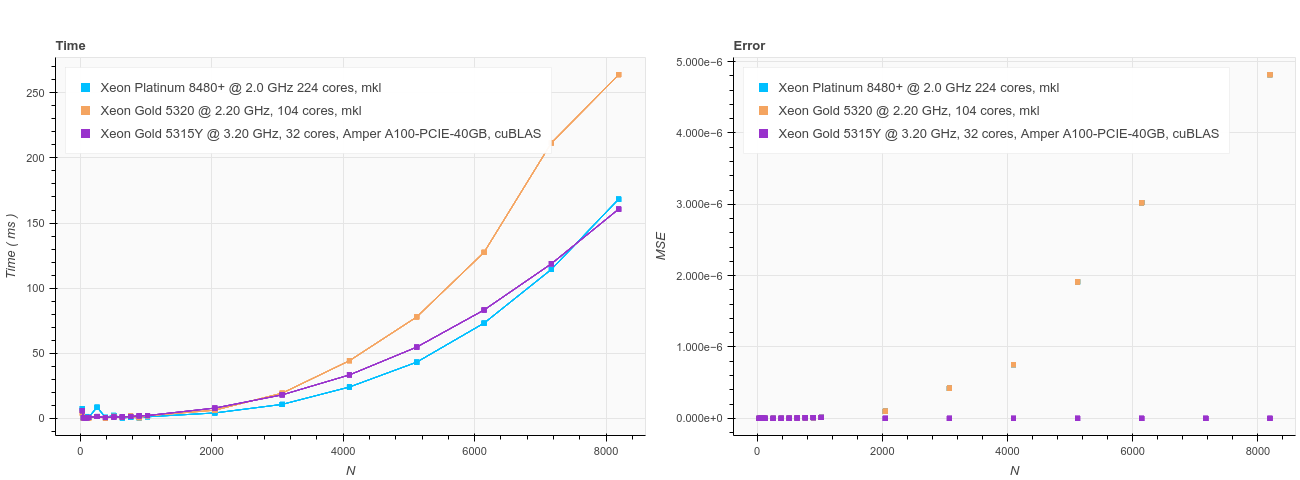}
\caption{Comparison between the best CPU and GPU execution times. Intel Xeon Platinum 8480+, Intel Xeon Gold 5320, and NVIDIA A100-PCIE-40GB GPU. MKL Vs. cuBLAS} \label{fig5}
\end{figure}

\begin{figure}
\includegraphics[width=\textwidth]{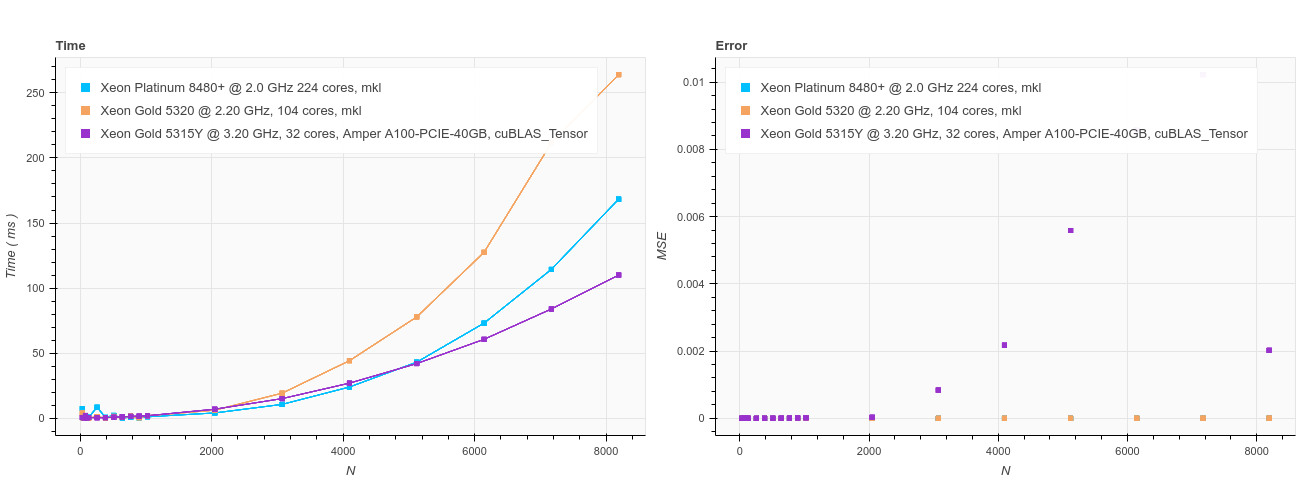}
\caption{Comparison between the best CPU and GPU execution times. Intel Xeon Platinum 8480+, Intel Xeon Gold 5320, and NVIDIA A100-PCIE-40GB GPU. MKL Vs. cuBLAS with Tensor Cores} \label{fig6}
\end{figure}

Figure \ref{fig5} compares the two best processors found and the NVIDIA A100. It is noted that the 104 cores at 2.2 GHz of the two Intel Xeon 5320 processors are far from the results of the NVIDIA A100, but the 224 cores at 2.0 GHz of the two Intel Xeon Platinum 8489+ processors with MKL present better time results than the NVIDIA A100 with cuBLAS. However, it should be noted that there was a more significant loss of accuracy in MKL. Using the Tensor Cores in cuBLAS increases the performance of the NVIDIA A100, which outperforms Intel's fourth-generation Platinum processor. In this case, the accuracy problem is reversed, and cuBLAS with Tensor Cores has a higher accuracy error than MKL, as shown in Figure \ref{fig6}.

\subsection{Power Consumption}
This section shows the power consumptions recorded through PAPI and PERF of the different CPUs and GPUs recorded in this work. The consumption for the Intel Xeon Platinum 8480+ processor could not be recorded because the PAPI RAPL events do not support this generation of processors, and only the total consumption can be recorded by PERF.

\begin{figure}
\includegraphics[width=\textwidth]{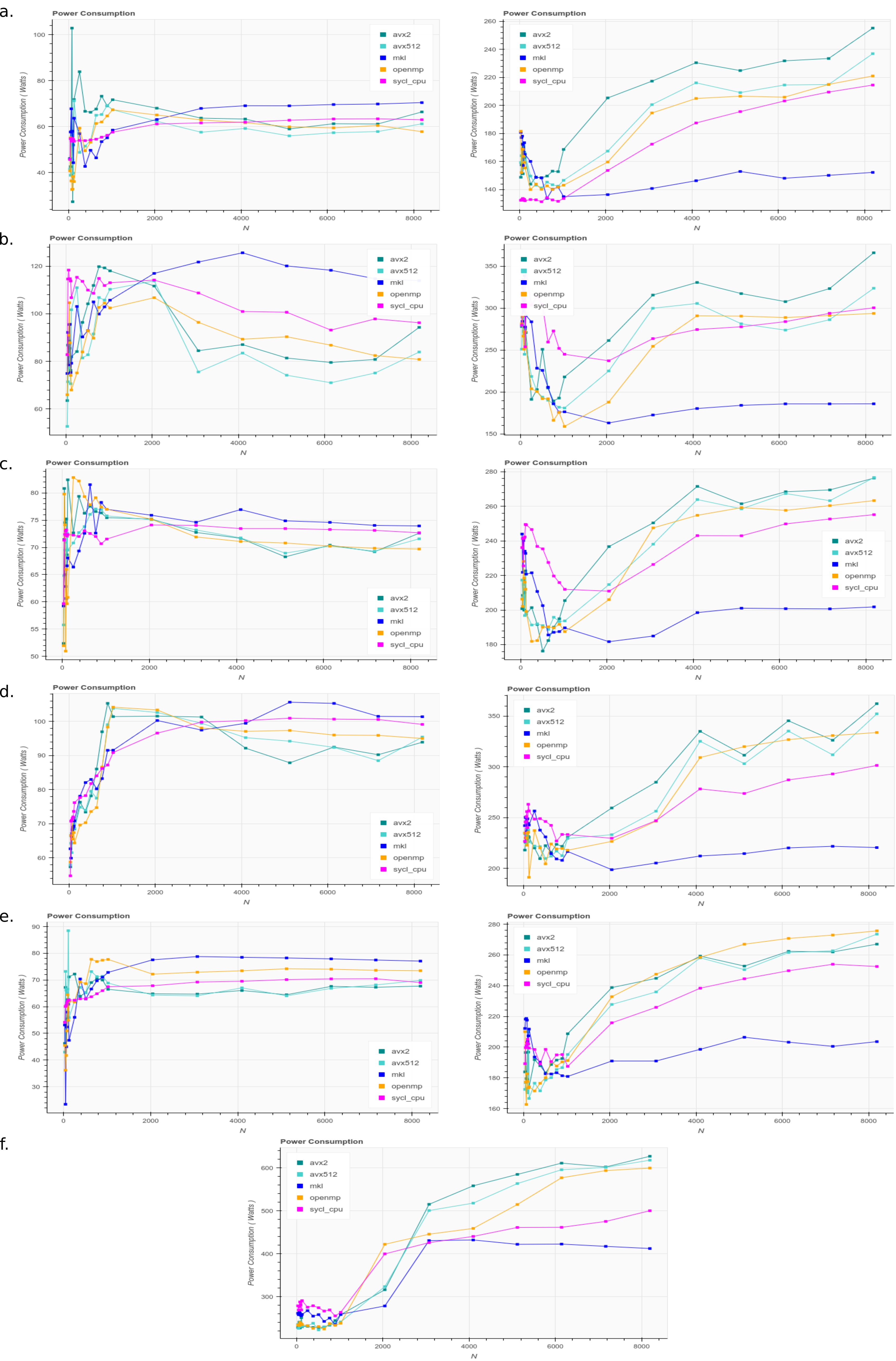}
\caption{The power consumption results obtained with PAPI are shown on the left, while the PERF results are on the right. a. Intel Xeon Gold 6126 b. Intel Xeon Gold 6254 c. Intel Xeon Silver 4314 d. Intel Xeon Gold 5320 e. Intel Xeon Gold 5315Y f. Intel Xeon Platinum 8480+} \label{fig7}
\end{figure}

Figure \ref{fig7} shows the power consumption recorded by both PAPI (left side) and PERF (right side). The power consumption recorded by PAPI of the matrix multiplication function for MKL is the highest compared to the other methods evaluated. However, the opposite behavior is presented when looking at what was recorded by PERF because it considers the power consumption of all code written, including array initialization, array loads from disk to memory, and others. For SYCL, it is observed that it presents less energy consumption in the function than MKL according to PAPI measurements but greater consumption with PERF measurements.

The Intel Xeon Platinum 8480+ processor registered the highest power consumption in PERF compared to the other processors. Finally, since we do not have metrics in PAPI for this processor, we cannot determine the exact power consumption of MKL in the function, so we will not compare it with the others.

\begin{figure}
\includegraphics[width=\textwidth]{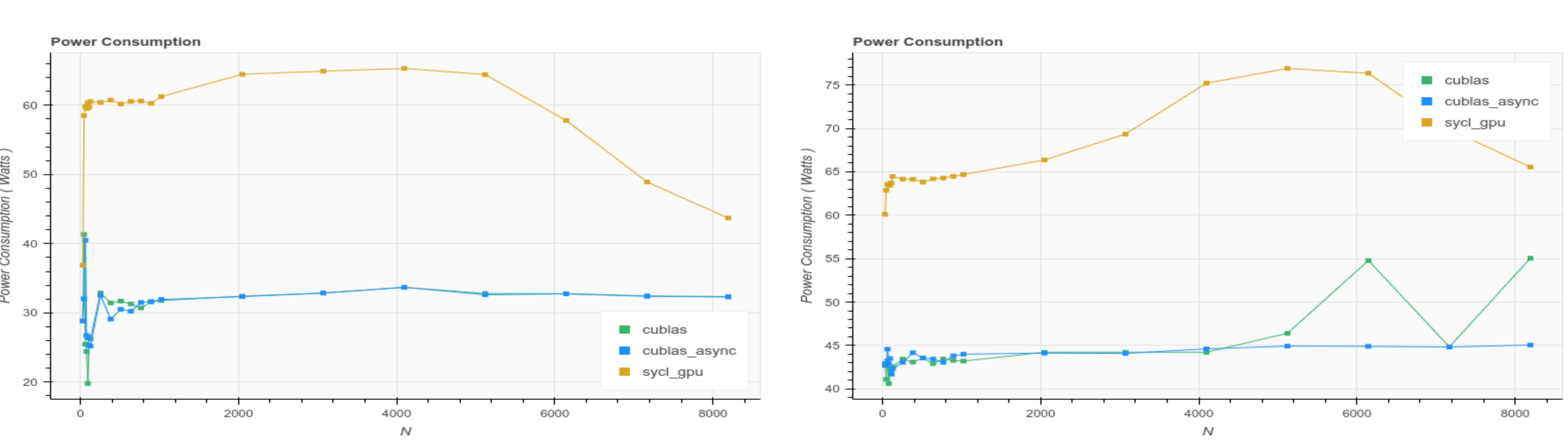}
\caption{Power consumption of NVIDIA V100 (left side) and NVIDIA A100 (right side). Comparison between SYCL, cuBLAS and cuBLAS with Tensor Cores.} \label{fig8}
\end{figure}

Figure \ref{fig8} shows the power consumption by cuBLAS and SYCL on GPUs. These measurements include CPU, memory, and GPU power consumption. It is noted that both the NVIDIA V100 and the A100 have similar power consumption for the cuBLAS library. On the other hand, SYCL had the highest consumption. However, there is a drop in consumption for matrices with N greater than 5120 due to the design of SYCL itself, which seeks energy efficiency in addition to portability, as mentioned in the work of Faqir-Rhazoui\cite{Faqir-Rhazoui}.

\begin{figure}
\includegraphics[scale=1]{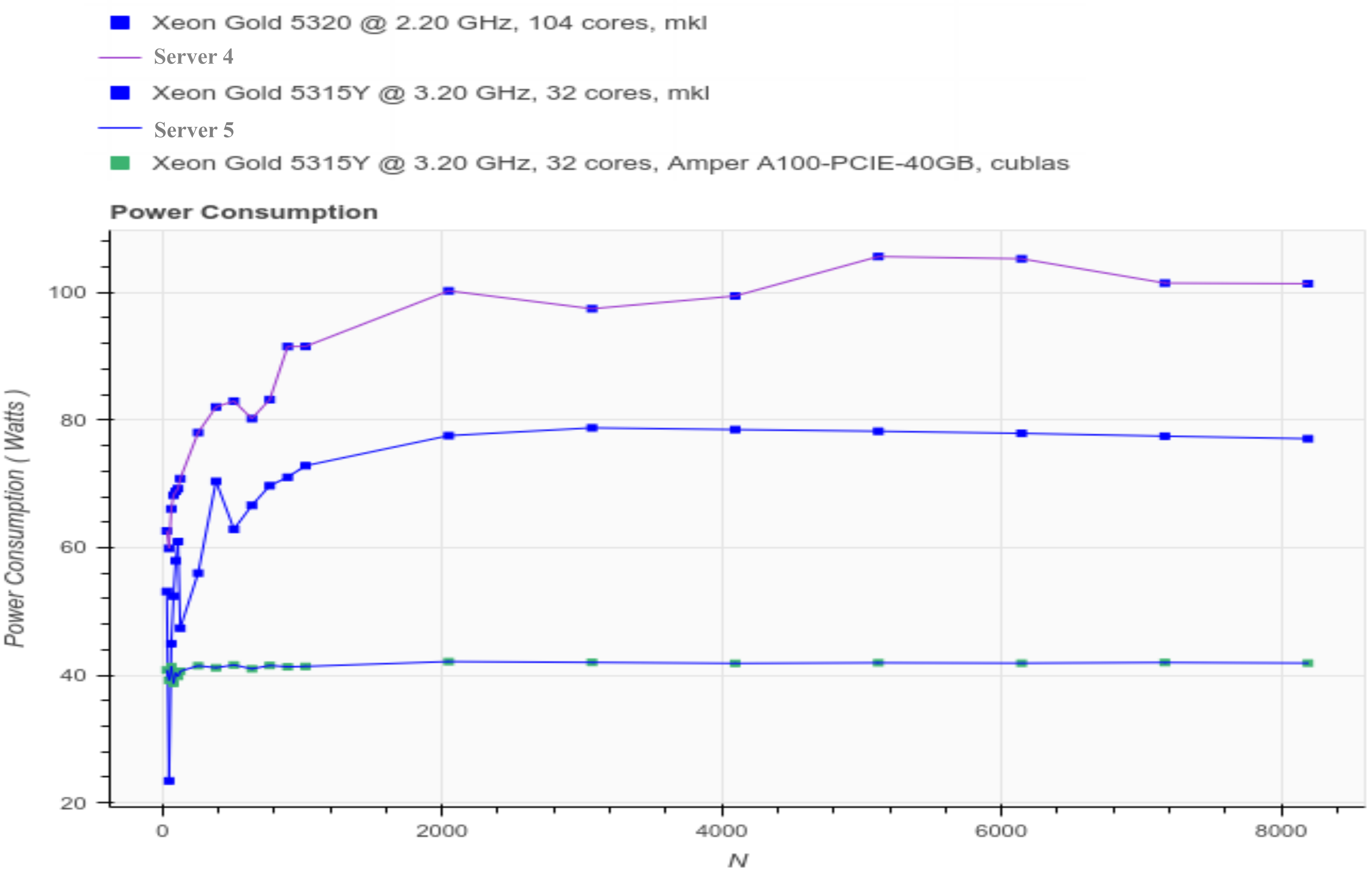}
\centering
\caption{Comparison of power consumption between the Intel Xeon Gold 5320, Intel Xeon Gold 5315Y processors with the MKL library and the NVIDIA A100 GPU with the cuBLAS library.} \label{fig9}
\end{figure}

Finally, Figure \ref{fig9} contrasts MKL's power consumption on the Intel Xeon Gold 5320 processor, the Intel Xeon Gold 5315Y processor, and cuBLAS on the NVIDIA A100. The increase in the number of cores and frequency causes a significant increase in power consumption. On the other hand, the NVIDIA A100 has almost constant power consumption for both small and large matrices because the workloads sent to the GPU fail to exert more significant stress on the device, unlike the CPU. 

The evaluation codes and results can be found in the following repository: \url{https://github.com/alejandrotorresn/MatMul} 
\section{Conclusion}
The main objective of this work was to determine if it is possible to match the processing capabilities of the GPU in terms of matrix multiplication with the CPU. As seen in the results presented in Figure 5, the Intel Xeon Platinum 8480+ processor with MKL managed to outperform the NVIDIA A100 when cuBLAS did not use Tensor Cores. However, the MSE is higher in this case for MKL, whereas in the case of Figure 6, where Tensor Cores are used, the NVIDIA A100 outperformed the processor only for arrays larger than 5120x5120 and with a much higher MSE than MKL. Therefore, based on the results presented, it is possible to match and even exceed GPU performance in matrix multiplication operations with at least fourth-generation Intel CPUs and similar characteristics to the evaluated processor.

Concerning power consumption, specific consumption could not be determined with PAPI on this processor. However, the results shown in Figure 7 show that the power consumption values recorded with PERF are about twice those recorded by PAPI for the MKL library. Therefore, the MKL consumption in this processor would be at least 200 Watts, much higher than the 45 to 55 Watts recorded for the NVIDIA A100 using cuBLAS.

Finally, the results show the feasibility of sending workloads of the backpropagation algorithm to the CPU, obtaining a performance similar to or superior to the GPU but with a significant increase in power consumption. It should be clarified that this is within the framework of the architectures evaluated; however, for the date of this study, Intel has launched its fifth generation of processors, and NVIDIA is already marketing its  Hopper architecture, so it is advisable to carry out a new evaluation. On the other hand, evaluating the impact of MSE concerning time in successive multiplications is recommended, as is the case with the training of neural networks.

\begin{credits}
\subsubsection{\ackname} Experiments presented in this paper were carried out using the Grid'5000 testbed, supported by a scientific interest group hosted by INRIA and including CNRS, RENATER and several Universities as well as other organizations in France\footnote{https://www.grid5000.fr}, the advanced computing platforms of the  High Performance and Scientific Computing Center at Universidad Industrial de Santander (SC3UIS)\footnote{http://www.sc3.uis.edu.co} and Universidad de Cartagena in Colombia.

\end{credits}
%
%
%

\begin{thebibliography}{8}

\bibitem{Cussen2023matrix}
Cussen, D., Ullman, J.D.: Matrix Multiplication Using Only Addition (2023). https://arxiv.org/abs/2307.01415v1

\bibitem{Rumelhart1986bp}
Rumelhart, D., Hinton, G. Williams, R. Learning representations by back-propagating errors. Nature \textbf{323}, 533--536 (1986). https://doi.org/10.1038/323533a0

\bibitem{Strassen1969Gaussian}
Strassen, V.: Gaussian Elimination is not Optimal. Numerische Mathematik \textbf{13}, 354-356 (1969). http://eudml.org/doc/131927

\bibitem{Alman2021FasterMM}
Alman, J., Williams, V.V.: A refined laser method and faster matrix multiplication. Proceedings of the Thirty-Second Annual ACM-SIAM Symposium on Discrete Algorithms, 522–-539 (2021). https://doi.org/10.1137/1.9781611976465.3

\bibitem{Qin2020SparceGEMM}
Qin, E. et al.: SIGMA: A Sparse and Irregular GEMM Accelerator with Flexible Interconnects for DNN Training. 2020 IEEE International Symposium HPCA, San Diego, CA, USA, 58--70 (2020). https://doi.org/10.1109/HPCA47549.2020.00015

\bibitem{Goto2008AnatomyMM}
Goto, K., Geijn, R. A. van de: Anatomy of high-performance matrix multiplication. Association for Computing Machinery, \textbf{34}{3}, 1–-25 (2008). https://doi.org/10.1145/1356052.1356053

\bibitem{Kuzma2023FastMM}
Kuzma B, Korostelev I, de Carvalho JPL, et al: Fast matrix multiplication via compiler-only layered data reorganization and intrinsic lowering. Softw: Pract Exper \textbf{53}{9}, 1793–-1814 (2023). https://doi.org/10.1002/spe.3214

\bibitem{Baratta2022PerfMat}
Baratta, I., Richardson, C., Wells, G. Performance analysis of matrix-free conjugate gradient kernels using SYCL. International Workshop on OpenCL \textbf{14}, 1-–10 (2022). https://doi.org/10.1145/3529538.3529993

\bibitem{Khalilov2021PerfCUDA}
Khalilov, M., Timoveev, A.: Performance analysis of CUDA, OpenACC and OpenMP programming models on TESLA V100 GPU. Journal of Physics: Conference Series \textbf{1740} (2021). https://dx.doi.org/10.1088/1742-6596/1740/1/012056

\bibitem{Krainiuk2021oneAPI}
Krainiuk, M., Goli, M., Pascuzzi, V. R.: oneAPI Open-Source Math Library Interface. International Workshop on Performance, Portability and Productivity in HPC (P3HPC), St. Louis, MO, USA. 22--32 (2021). doi: 10.1109/P3HPC54578.2021.00006.

\bibitem{Markidis2018TensorProg}
Markidis, S., Chien, S., Laure, E., Peng, I., Vetter, J. : NVIDIA Tensor Core Programmability, Performance and Precision. 2018 IEEE IPDPSW, Vancouver, BC, Canada. 522-531 (2018). https://doi.ieeecomputersociety.org/10.1109/IPDPSW.2018.00091

\bibitem{Yan2020DemysTensor}
Yan, D., Wang, W., Chu, X.: Demystifying Tensor Cores to Optimize Half-Precision Matrix Multiply. 2020 IEEE IPDPS, New Orleans, LA, USA. 634-643 (2020). https://doi.org/10.1109/IPDPS47924.2020.00071

\bibitem{Reddy2021PergGPU}
Reddy, G.K., Vaidya, R., Barve, M.: Performance Study of GPU applications using SYCL and CUDA on Tesla V100 GPU. 2021 IEEE High Performance Extreme Computing Conference (HPEC), Waltham, MA, USA. 1-7 (2021), https://doi.org/10.1109/HPEC49654.2021.9622813

\bibitem{daSilva2016CompSYCL}
da Silva, H.C., Pisani, F., Borin, E.: A Comparative Study of SYCL, OpenCL, and OpenMP. 2016 International SBAC-PADW, Los Angeles, CA, USA. 61-66 (2026). https://doi.org/10.1109/SBAC-PADW.2016.19

\bibitem{Reguly2019PortKernels}
Reguly, I.Z.: Performance Portability of Multi-Material Kernels. 2019 IEEE/ACM International Workshop on Performance, Portability and Productivity in HPC (P3HPC), Denver, CO, USA. 26-35 (2019). https://doi.org/10.1109/P3HPC49587.2019.00008

\bibitem{Jo02019PerfPort}
Rangel, E.M., Pennycook, S.J., Pope, A., Frontiere, N., Ma, Z., Madananth, V.: A Performance-Portable SYCL Implementation of CRK-HACC for Exascale. Proceedings of the SC '23 Workshops of The International Conference on High Performance Computing, Network, Storage, and Analysis, 1114–-1125 (2023). https://doi.org/10.1145/3624062.3624187

\bibitem{Hammond2019CompKokkos}
Hammond, J., Kinsner, M., Brodman, J.: A comparative analysis of Kokkos and SYCL as heterogeneous, parallel programming models for C++ applications. IWOCL'19: Proceedings of the International Workshop on OpenCL, (2019). https://doi.org/10.1145/3318170.3318193

\bibitem{Faqir-Rhazoui}
Faqir-Rhazoui, Y., García, C. SYCL in the edge: performance and energy evaluation for heterogeneous acceleration. J Supercomput (2024). https://doi.org/10.1007/s11227-024-05957-6

\end{thebibliography}
%

\end{document}